\documentclass{article}
\usepackage{spconf,amsmath,graphicx}
\usepackage{hyperref}
\usepackage{mathtools}
\usepackage{tcolorbox}
\usepackage{color}
\usepackage{theorem}
\usepackage{times,amsmath,epsfig}
\usepackage{amssymb}
\usepackage{subfigure}
\usepackage{cite}
\usepackage{cases}
\usepackage{siunitx}
\usepackage{algorithmic}
\usepackage{algorithm}
\usepackage{needspace}

\usepackage{tikz}
\usetikzlibrary{shapes,arrows}
\input{mysymbol.sty}
\tikzstyle{phantom vertex} = [ ellipse, 
                               anchor = center, 
                               minimum height = 1*\unit, 
                               minimum width  = 1*\unit,
                               inner sep=0pt,
                               anchor=center]
\tikzstyle{red vertex}   = [black, fill = red!20,   phantom vertex, draw]
\tikzstyle{black vertex} = [black, fill = black!20, phantom vertex, draw]
\tikzstyle{blue vertex}  = [black, fill = blue!20,  phantom vertex, draw]
\tikzstyle{green vertex} = [black, fill = green!20,  phantom vertex, draw]
\tikzstyle{yellow vertex} = [black, fill = yellow!20,  phantom vertex, draw]
\tikzstyle{cyan vertex} = [black, fill = cyan!20,  phantom vertex, draw]
\tikzstyle{vertex}       = [draw, phantom vertex]

\tikzstyle{point} = [ellipse, inner sep=0pt, draw, fill=white, anchor = center,
                     minimum height = 0.05*\unit, minimum width  = 0.05*\unit]

\newtheorem{myproposition}{\bf Proposition}

\title{Efficient power allocation using graph neural networks and deep algorithm unfolding}
\name{Arindam Chowdhury$^\star$, Gunjan Verma$^\dag$, Chirag Rao$^\dag$, Ananthram Swami$^\dag$, and Santiago Segarra$^\star$
\thanks{Research was sponsored by the Army Research Office and was accomplished under Cooperative Agreement Number W911NF-19-2-0269. 
		The views and conclusions contained in this document are those of the authors and should not be interpreted as representing the official policies, either expressed or implied, of the Army Research Office or the U.S. Government. 
		The U.S. Government is authorized to reproduce and distribute reprints for Government purposes notwithstanding any copyright notation herein.
		\newline
		Emails:  \{ac131, segarra\}@rice.edu, \{gunjan.verma.civ, chirag.r.rao.civ, ananthram.swami.civ\}@mail.mil. }}
\address{$^\star$Rice University, USA  \hspace{1cm} $^\dag$US Army’s CCDC Army Research Laboratory, USA}
\begin{document}
\ninept
\renewcommand{\baselinestretch}{0.95}
\maketitle
\begin{abstract}
    We study the problem of optimal power allocation in a single-hop ad hoc wireless network.
	In solving this problem, we propose a hybrid neural architecture inspired by the algorithmic unfolding of the iterative weighted minimum mean squared error (WMMSE) method, that we denote as unfolded WMMSE (UWMMSE). 
	The learnable weights within UWMMSE are parameterized using graph neural networks (GNNs), where the time-varying underlying graphs are given by the fading interference coefficients in the wireless network. These GNNs are trained through a gradient descent approach based on multiple instances of the power allocation problem. Once trained, UWMMSE achieves performance comparable to that of WMMSE while significantly reducing the computational complexity. This phenomenon is illustrated through numerical experiments along with the robustness and generalization to wireless networks of different densities and sizes.
\end{abstract}
\section{Introduction}
\label{sec:intro}

Power and bandwidth are fundamental resources that determine the effective capacity of a wireless channel~\cite{shannon1948mathematical}. 
Hence, optimal allocation of these resources under randomly varying channel characteristics and user demands is essential for the smooth operation of wireless systems. 
In particular, power allocation in a wireless ad hoc network is crucial to mitigate multi-user interference, one of the main performance-limiting factors.
Mathematically, power allocation can be formulated as the problem of optimizing a certain system-level utility function (such as sum rate or harmonic rate) subject to resource budget constraints.
Despite the remarkable success of this paradigm~\cite{boche2011characterization}, many of the associated optimization problems are non-convex and NP-hard~\cite{luo_2008_dynamic,razaviyayn2013linear}.
For the canonical case of sum-rate maximization, the most commonly used algorithm is weighted minimum mean squared error (WMMSE) minimization~\cite{shi2011iteratively}. 
In spite of being close to optimal in practice, it has  high computational complexity and slow convergence. 
This has led the research community to look for faster, data-driven solutions for power allocation.

Deep learning based methods have emerged as promising alternatives to classical methods for resource allocation in recent years~\cite{qin2019deep}. 
In \emph{supervised} settings, deep neural networks learn to mimic established classical methods from solved instances~\cite{sun2018learning}. 
In contrast, some works have followed an \emph{unsupervised} approach, where one parameterizes the power allocation function using a neural network and directly employs the optimization objective as a loss function, bypassing the need for solved problem instances~\cite{lee2018deep,eisen2019learning}. While such a procedure is computationally simple, no prior knowledge (e.g., based on classical optimization techniques) is leveraged to inform the algorithm's architectural or hyper-parameter choices.

We advocate a third direction that is \emph{unsupervised} as no solved instances of the power allocation problem are needed for training, while \emph{imitating} classical methods by incorporating part of their structure into the layered architecture of the neural network.
The goal is to leverage the theoretical models developed with expert knowledge and achieve near-optimal performance with significantly reduced execution time.
To accomplish this, we follow the paradigm of algorithm unfolding~\cite{gregor2010learning,monga2019algorithm}.
The idea is to unfold the iterations of a classical iterative algorithm as a cascade of layers, where each layer has the same update structure as the original algorithm but the parameters can be learned from data. 
A standard method of unfolding is to parameterize the function of interest using multi-layer perceptrons (MLPs) or convolutional neural networks (CNNs)~\cite{guo2019structure}. 
However, MLPs and CNNs are not quite suitable for problems in wireless communication. 
In particular, their performance degrades dramatically when the network size becomes large since MLPs and CNNs cannot exploit the underlying network topology. 
We adopt an alternative direction~\cite{eisen2020optimal, nakashima2019deep, shen2019graph}, where graph neural networks (GNNs) \cite{bruna2013spectral,kipf2016semi,defferrard2016convolutional,li2017diffusion,gama2018convolutional,roddenberry_2019_hodgenet,yang_2018_enhancing} are used to parameterize the power allocation function, thus leveraging the natural representation of wireless networks as graphs. GNNs utilize the structural relationships between nodes to locally process instantaneous channel state information.
In this context, we propose an unfolded weighted minimum mean squared error (UWMMSE) method, which is to the best of our knowledge the first GNN-based deep unfolded architecture based on the iterative WMMSE algorithm. UWMMSE simultaneously achieves state-of-the-art performance in utility maximization and computational efficiency for power allocation in wireless networks.

\medskip\noindent\textbf{Contribution.} 
The contributions of this paper are twofold:\\
i) We propose an unfolded version of WMMSE for power allocation in wireless networks, where the learnable modules are parameterized through GNNs.\\
ii) We empirically illustrate the performance of the proposed method, compare it with state-of-the-art alternatives, and demonstrate its generalization to networks of unseen sizes and densities.

%%%%%%%%%%%%%%%%%%%%%%%%%%%%%%%%%%%%%%%%%%%%
\section{System model and problem formulation}\label{S:Modeling}
%%%%%%%%%%%%%%%%%%%%%%%%%%%%%%%%%%%%%%%%%%%%

We consider a single-hop ad hoc interference network having $M$ distinct single-antenna transceiver pairs. Transmitters are denoted by $i$ and the $i$-th transmitter is associated with a single receiver denoted by $r(i)$ for $i \in \{1, \ldots, M\}$.  
Further, denoting the signal transmitted by $i$ as $x_i \in \mathbb{R}$, the received signal $y_i \in \mathbb{R}$ at $r(i)$ is given by
\begin{equation}\label{E:trans_model}
     y_i = h_{ii}x_i + \sum_{\substack{j=1 \, | \,  j\neq i}}^M h_{ij}x_j + n_i,
\end{equation}
where $h_{ii} \in \mathbb{R}$ is the channel between the $i$-th transceiver pair, $h_{ij} \in \mathbb{R}$ for $i \neq j$ represents the interference between transmitter $j$ and receiver $r(i)$, and $n_i \sim \ccalN(0, \sigma^2)$ % for $\sigma > 0$   
represents the additive channel noise. We consider time-varying channel states $h_{ij}(t)$, which are stored in a channel-state matrix $\bbH(t) \in \reals^{M \times M}$ where $[\bbH(t)]_{ij} = h_{ij}(t)$.
We can interpret $\bbH(t)$ as the weighted adjacency matrix of a directed graph with $M$ nodes, where node $i$ represents the $i$-th transceiver pair.
The instantaneous data rate $c_i$ achievable at receiver $r(i)$ is given by Shannon's capacity theorem%as a function of the signal-to-interference-plus-noise-ratio (SINR), 
\begin{equation}\label{E:data_rate}
    c_i(\bbp(t), \bbH(t)) = \log_2 \left( 1 + \frac{ h_{ii}^2 (t) p_i(t)}{\sigma^2 + \sum_{j\neq i} h_{ij}^2(t) p_j(t)} \right),
\end{equation}
where $p_i(t) \ge 0$ % \in \mathbb{R}_+$
is the power allocated to transmitter $i$ at time $t$ and $\bbp(t) = [p_1(t), \ldots, p_M(t)]^\top$.
The objective is to determine the instantaneous power allocation vector $\bbp(t)$ that maximizes a network utility that is a function of data rates $c_i$. 
Omitting the explicit dependence on $t$ to simplify notation, we formalize the standard sum-rate maximization problem under power constraints as follows
\begin{align}\label{E:optimization_problem}
		 \max_{\bbp} \,\, \sum_{i=1}^M c_i(\bbp, \bbH) \qquad   \text{s.t.} \,\,\, 0 \leq p_i \leq p_{\max}, \,\, \text{for all} \,\, i,
\end{align}
where $p_{\max}$ denotes the maximum available power at every transmitter.
Though seemingly simple, this optimization problem has been shown to be NP-hard~\cite{luo_2008_dynamic, hong_2014_signal}.
In this work, we aim to achieve an \emph{effective} and \emph{efficient} solution to~\eqref{E:optimization_problem}, where $\bbH$ is drawn from an \emph{accessible} distribution $\ccalH$. Here, by \emph{effective} we mean a solution that achieves  performance close to that of a near-optimal classical approach, while \emph{accessible} distribution means that either $\ccalH$ is known or we can easily sample from it.

Classical approaches focus on finding approximate solutions for a single instance of~\eqref{E:optimization_problem} for an arbitrary $\bbH$, and then repeat this operation to recompute the power allocation in successive time instants. 
Given that in practice we are interested in solving several instances of~\eqref{E:optimization_problem} across time, a learning-based body of work has gained traction in the past years~\cite{eisen2020optimal,sun2018learning}.
In a nutshell, based on many channel state instances, the idea is to learn a map (i.e., a function approximator) between the channel state matrix $\bbH$ and the corresponding (approximate) optimal power allocation $\bbp$. 
Unlike common neural-network models whose inputs and outputs are often of fixed dimension, the dimension of $\bbH$ is not necessarily fixed a priori and can indeed change as nodes enter and exit the wireless network. This fact combined with the inherent topological structure present in $\bbH$ motivates the use of graph neural networks.

Our goal is to combine the advantages of the classical and learning-based paradigms by
leveraging the \emph{approximate} and \emph{interpretable} solution provided by the classical WMMSE \cite{shi2011iteratively} method while enhancing it with the computational run-time \emph{efficiency} of trained machine learning models. 
We pursue this synergistic combination under the paradigm of algorithm unfolding, as presented next.

\section{Algorithm unfolding for power allocation}\label{S:uwmmse}

%%%%%   F I G U R E  %%%%%%%%%%%%%%%%%%%%%
\begin{figure}[t]
	\centering
	\includegraphics[width=\linewidth]{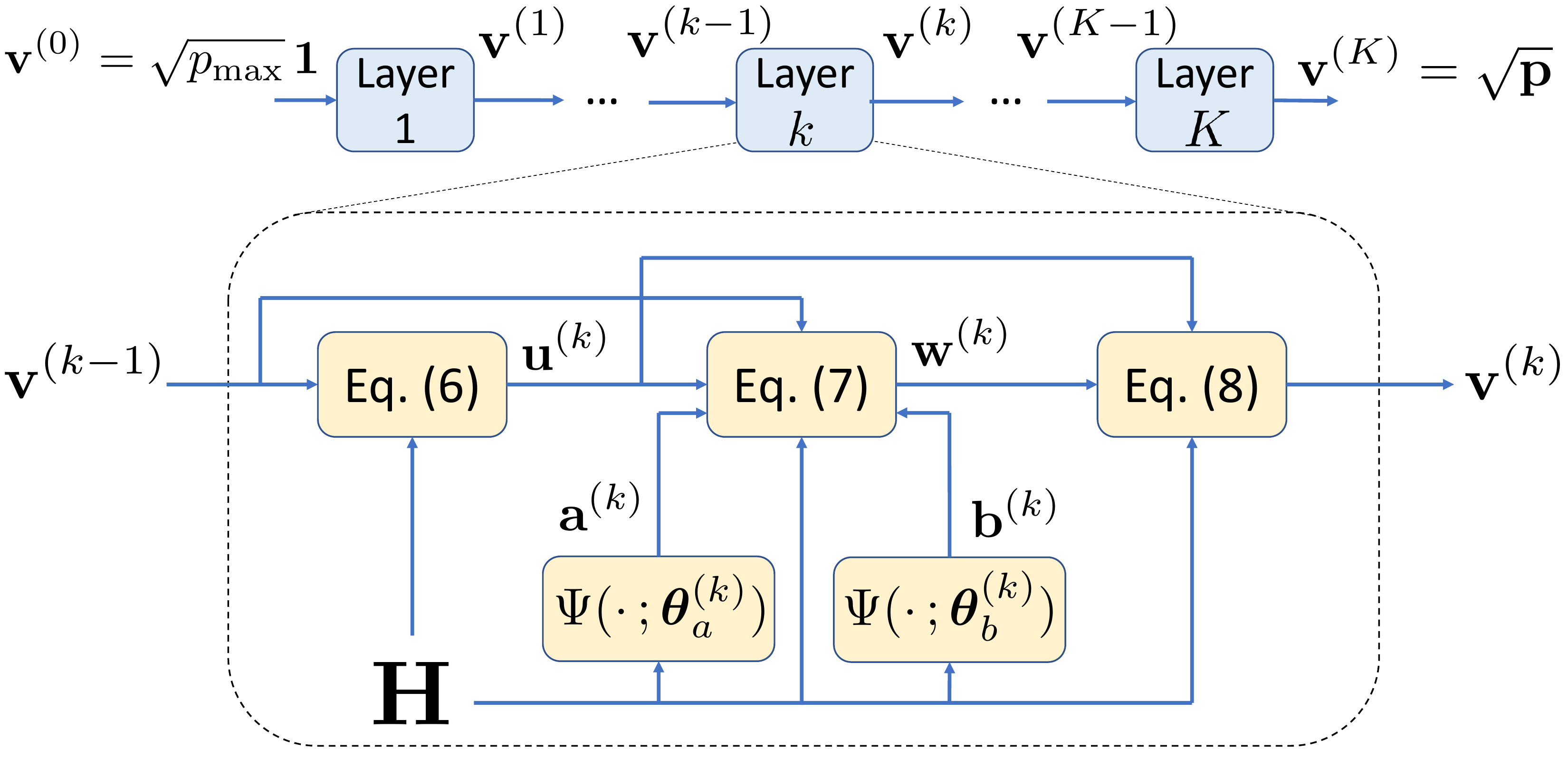}
	\vspace{-0.65cm}
	\caption{\small Schematic diagram of the proposed UWMMSE algorithm.
	A generic intermediate layer $k$ is detailed.}
	\label{F:unr}
\end{figure}
%%%%%%%%%%%%%%%%%%%%%%%%%%%%%%%%%%%%%%%

Algorithm unfolding~\cite{balatsoukas2019deep,liu2019deep,monga2019algorithm,farsad2020data} refers to the general notion of building a problem-specific neural architecture with layers inspired by an iterative solution to the same problem. 
In this paper, we develop and evaluate a way to unfold the classical WMMSE~\cite{shi2011iteratively} method using graph neural networks for power allocation in wireless networks.

We start by introducing the basics of WMMSE.
Essential for this classical algorithm is to reformulate~\eqref{E:optimization_problem} as
\begin{align}\label{E:problem_reformulation}
&\min_{\mathbf{w,u,v}} \sum_{i=1}^M (w_i e_i - \log w_i ),\\
& \text{s.t.} \,\,\,\,\, e_i = (1-u_i h_{ii} v_i)^2 + \sigma^2 u_i^2 + \sum_{i \neq j} u_i^2 h_{ij}^2 v_j^2, \,\,\,\, v^2_i \leq p_{\max},\nonumber
\end{align}
where the constraints are repeated for all $i$ and $e_i$ computes the mean-square error of the signal at node $i$ under the assumption that the transmitted signal $x_i$ is independent of the noise $n_i$ [cf.~\eqref{E:trans_model}]. 

We say that~\eqref{E:problem_reformulation} is equivalent to~\eqref{E:optimization_problem} because it can be shown~\cite[Thm. 3]{shi2011iteratively} that the optimal solution $\{\bbw^*, \bbu^*, \bbv^*\}$ of the former and that of the latter $\bbp^*$ are related as $\sqrt{\bbp^*} = \bbv^*$, where the square root is applied elementwise.
While~\eqref{E:problem_reformulation} is non-convex, it is convex in each variable when fixing the other two; this motivates block-coordinate descent, which provides closed-form updates rules to find a local minimum.
%Additionally, a solution to can be approximated via block-coordinate descent, since the objective is convex in each variable when the other two are fixed, providing closed-form update rules. 
Our unfolding architecture is inspired by these iteratively applied closed-form equations, which we augment with learnable parameters.

We propose to compute the allocated power as a function of the channel state matrix $\bbp = \Phi(\bbH; \bbTheta)$ through a layered architecture $\Phi$ with trainable weights $\bbTheta$. 
More precisely, setting $\bbv^{(0)} = \sqrt{p_{\max}} \, \mathbf{1}$, we have that for layers $k = 1, \ldots, K$, 
\begin{align}
\bba^{(k)} &= \Psi(\bbH; \bbtheta_a^{(k)}),  \qquad \bbb^{(k)}  = \Psi(\bbH;\bbtheta_b^{(k)}), \hspace{-10mm} \label{E:unfold_1}\\
u^{(k)}_i &= \frac{h_{ii}v^{(k-1)}_i}{\sigma^2 + \sum_j h_{ij}^2 {v^{(k-1)}_j} v^{(k-1)}_j}, \,\, &&\text{for all } i, \label{E:unfold_2}\\
w^{(k)}_i &= \frac{a_i^{(k)}}{1 - u^{(k)}_i h_{ii} v^{(k-1)}_i} + b_i^{(k)}, &&\text{for all } i, \label{E:unfold_3}\\
v^{(k)}_i &= \alpha \left( \frac{ u^{(k)}_i h_{ii} w^{(k)}_i}{\sum_j  h_{ji}^2 u^{(k)}_j u^{(k)}_j w^{(k)}_j}\right), &&\text{for all } i,\label{E:unfold_4}
\end{align}
and the output power is determined as $\bbp = \Phi(\bbH; \bbTheta) = (\bbv^{(K)})^2$, where the square is applied elementwise.
The non-linear function $\alpha(z) \coloneqq [z]_0^{\sqrt{p_{\max}}}$ in~\eqref{E:unfold_4} simply ensures that $v_i^{(k)} \in [0, \sqrt{p_{\max}}]$ by saturating the right-hand side of~\eqref{E:unfold_4} at these extreme values. 
This guarantees that the constraint in~\eqref{E:optimization_problem} is satisfied.
The trainable parameters $\bbTheta$ are given by the collection of $\bbtheta_a^{(k)}$ and $\bbtheta_b^{(k)}$ in~\eqref{E:unfold_1} for all $K$ layers.
Finally, the functions $\Psi$ parametrized by $\bbTheta$ in~\eqref{E:unfold_1} are chosen to be graph convolutional networks (GCNs)~\cite{kipf2016semi}. 
A schematic view of our proposed layered architecture is presented in Fig.~\ref{F:unr}.

In the proposed architecture, notice that each layer $k$ as described in \eqref{E:unfold_1}-\eqref{E:unfold_4} is characterized by five vectors $\bba^{(k)}, \bbb^{(k)}, \bbu^{(k)},$ $\bbw^{(k)}, \bbv^{(k)} \in \reals^M$. 
If $\bba^{(k)} = \mathbf{1}$ and $\bbb^{(k)} = \mathbf{0}$, expressions \eqref{E:unfold_2}-\eqref{E:unfold_4} correspond to the closed-form expressions for block coordinate descent on~\eqref{E:problem_reformulation}; see~\cite{shi2011iteratively} for details. 
In other words, for these values of $\bba^{(k)}$ and $\bbb^{(k)}$, our proposed architecture boils down to a truncated WMMSE with $K$ iterations. 
In this setting, $\bbu^{(k)}$ and $\bbv^{(k)}$ represent receiver and transmitter side variables, respectively, such that $u_i^{(k)}$ depends exclusively on the channel states \emph{into} receiver $r(i)$ whereas $v_i^{(k)}$ depends exclusively on the channel states \emph{out of} transmitter $i$. 

A major drawback of WMMSE lies in its high computational and time complexity. This complexity arises because WMMSE requires many iterations of the updates \eqref{E:unfold_2}-\eqref{E:unfold_4} for convergence. 
Hence, the objective of the learned variables $\bba^{(k)}$ and $\bbb^{(k)}$ is to accelerate this convergence while maintaining good performance.
Intuitively, if we learn a smarter update rule for $\bbw^{(k)}$ that accelerates its convergence, we can achieve good performance with only a few iterations of WMMSE. 
% In this sense, our proposed architecture is an unfolded version of the classical algorithm.
Notice that additional learning parameters and more sophisticated functional forms could be included in the updates~\eqref{E:unfold_2}-\eqref{E:unfold_4}. 
However, the learned affine transformation proposed in~\eqref{E:unfold_3} achieves good performance in practice (see Section~\ref{S:num_exp}) while being simple to implement.

Under the natural assumption that the weights $\bba^{(k)}$ and $\bbb^{(k)}$ should depend on the channel state $\bbH$, we advocate a learning-based method where this dependence is made explicit via the parametric functions $\Psi$ in~\eqref{E:unfold_1}. 
For fixed parameters $\bbTheta$, the allocated power for a channel state $\bbH$ is given by $\Phi(\bbH; \bbTheta)$ and results in a sum-rate utility of $\sum_{i=1}^M c_i(\Phi(\bbH; \bbTheta), \bbH)$. 
Hence, we define the loss function
\begin{equation}\label{E:loss_sgd}
\ell(\bbTheta) = - \mathbb{E}_{\bbH \sim \ccalH} \left[ \sum_{i=1}^M c_i(\Phi(\bbH; \bbTheta), \bbH) \right].
\end{equation}
Leveraging the facts that GCNs $\Psi(\cdot\, ; \bbtheta)$ in~\eqref{E:unfold_1} are differentiable with respect to $\bbtheta$ and that we have access to samples of $\ccalH$ (cf. Section~\ref{S:Modeling}), we seek to minimize~\eqref{E:loss_sgd} through stochastic gradient descent. 
Notice that we think of UWMMSE as an \emph{unsupervised method} in the sense that, for training, it requires access to samples of the channel state matrices $\bbH$ but does \emph{not} require access to the optimal power allocations (output values) associated with those channels.

Finally, note that the indexing of the $M$ transceiver pairs is arbitrary. 
Hence, any reasonable power allocation policy should be independent of these indices, which is formally encoded in the notion of permutation equivariance.
Given a function $f: \reals^{M \times M} \to \reals^M$, we say that $f$ is \emph{permutation equivariant} if $f(\bbPi \bbH \bbPi^\top) = \bbPi f(\bbH)$ for all $\bbH$ and all permutation matrices $\bbPi$. Naturally, the choice of $\Psi$ in~\eqref{E:unfold_1} affects the permutation equivariance of our proposed method, as we state next.\footnote{The proof, omitted due to space limitations, can be found in \cite{chowdhury2020unfolding}.}

\begin{myproposition}\label{P:equiv}
If $\Psi( \cdot\,; \bbtheta)$ in~\eqref{E:unfold_1} is permutation equivariant then the UWMMSE method $\Phi(\cdot\,; \bbTheta)$ is also permutation equivariant.
\end{myproposition}	

Given that GCNs, our choice for $\Psi$, are known to be permutation equivariant~\cite{kipf2016semi}, Proposition~\ref{P:equiv} indicates that UWMMSE is permutation equivariant, as desired.

%%%%%%%%%%%%%%%%%%%%%%%%%%%%%%%%%%%%%%%%%%%%
\section{Numerical experiments}\label{S:num_exp}

\begin{figure*}
	\centering
	\subfigure[]{
			\centering
			\includegraphics[width=0.31\textwidth,height=0.22\textwidth]{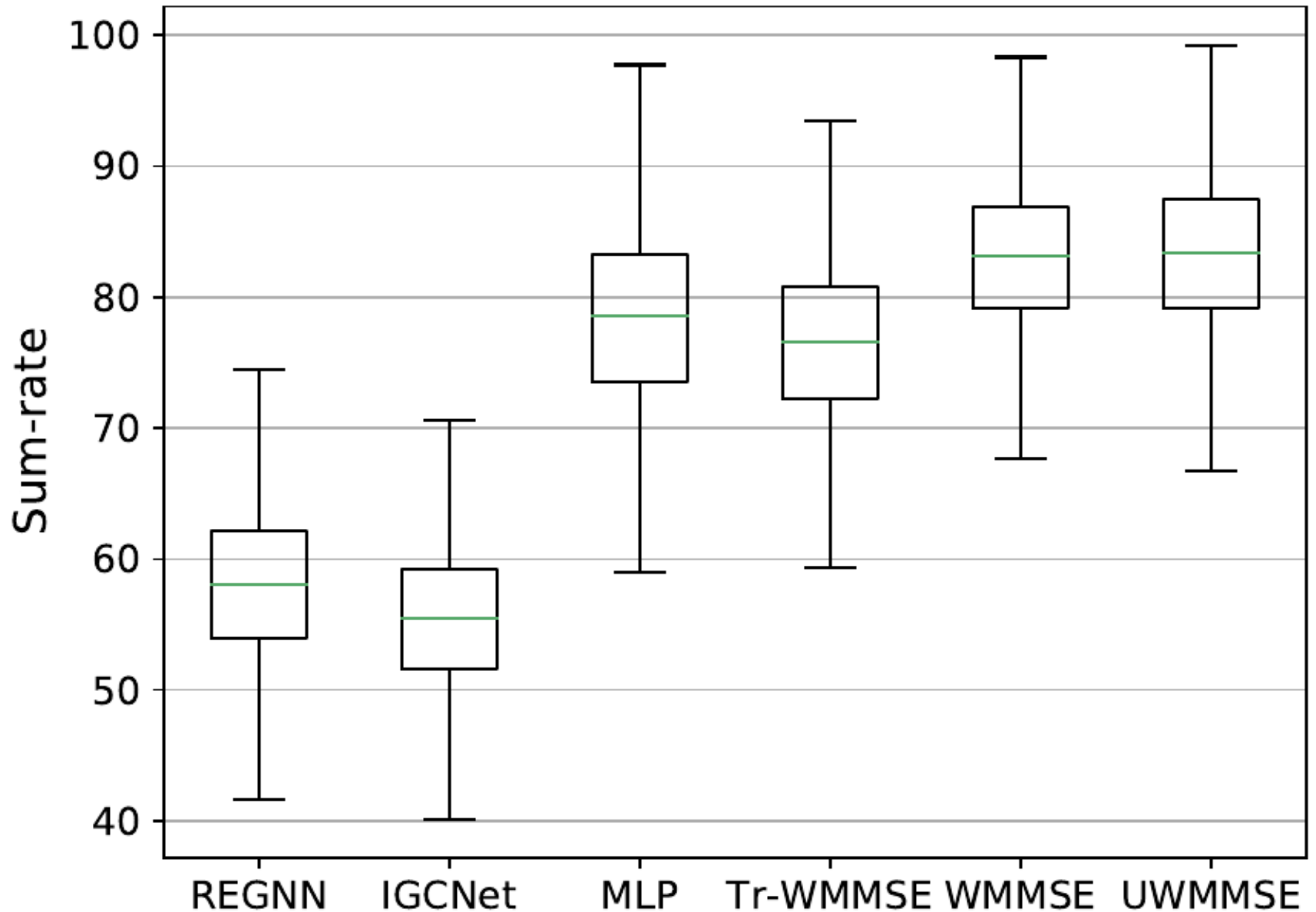}
			\label{Fig:performance_comparison_1}
		}	
	\subfigure[]{
			\centering
			\includegraphics[width=0.30\textwidth,height=0.22\textwidth]{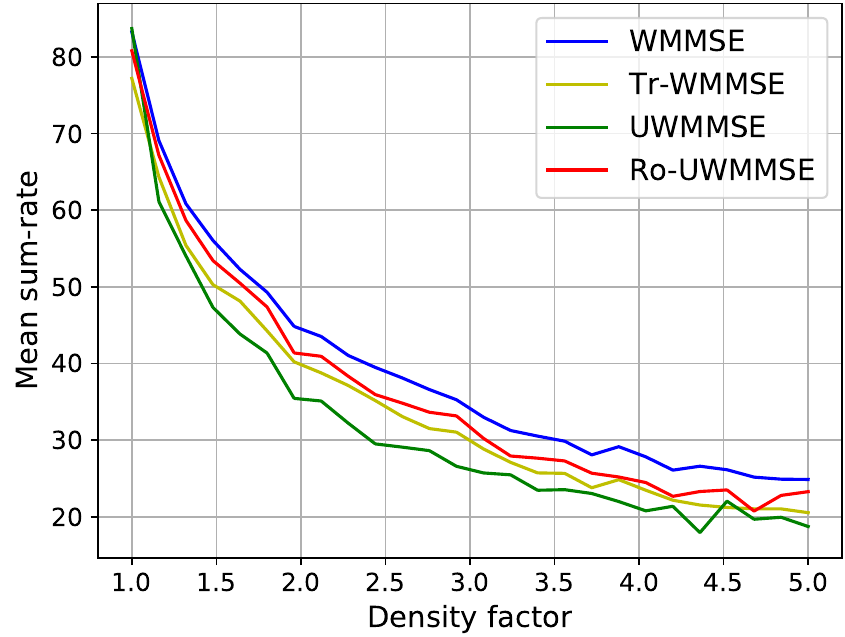}
			\label{Fig:performance_comparison_2}
		}	
	\subfigure[]{
			\centering
			\includegraphics[width=0.30\textwidth,height=0.22\textwidth]{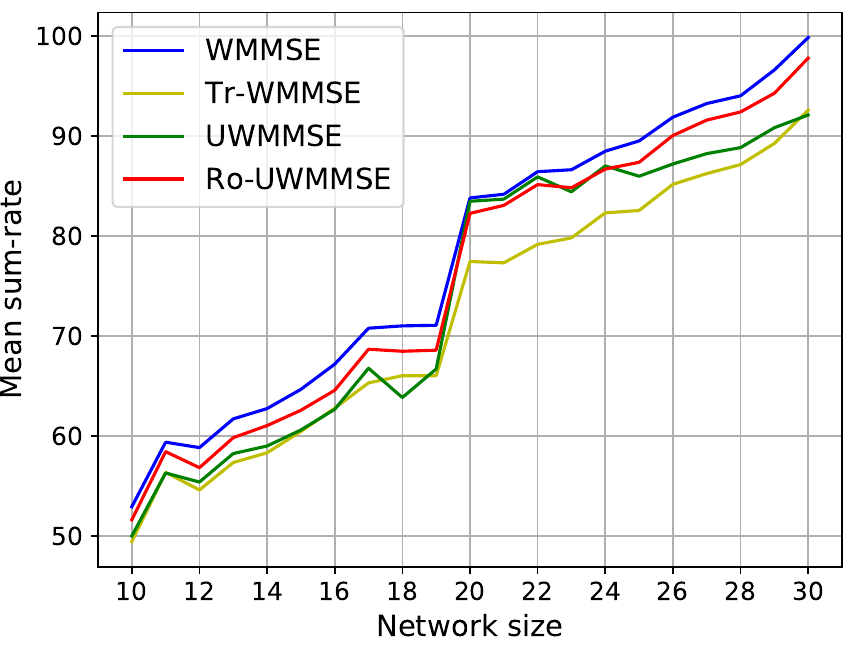}
			\label{Fig:performance_comparison_3}
		}
		\vspace{-0.4cm}
		\caption{\small {Performance and generalization analysis for UWMMSE. 
		(a)~Box plots of the achieved utilities for $6400$ randomly drawn channel state matrices.
		(b)~Mean sum-rate achieved as density factor $d$ varies in $[1.0,5.0]$, generating denser networks.
		(c)~Mean sum-rate achieved as network size $M$ varies in $\{10, \ldots,30\}$.}}
		\label{Fig:performance_comparison}
\end{figure*}

We simulate a \textit{Rayleigh fading} channel as a test-bed for our experiments.\footnote{Code to replicate the numerical experiments here presented can be found at \href{https://github.com/ArCho48/Unrolled-WMMSE.git}{https://github.com/archo48/unrolled-wmmse.git}.}
To that end, we construct a random geometric graph in two dimensions having $M$ transceiver pairs. 
First, each transmitter $i$ is dropped uniformly at random at locations $\mathbf{t}_i \in [-M, M]^2$. 
Then, its paired receiver $r(i)$ is dropped uniformly at random at location $\mathbf{r}_i \in [\bbt_i - \frac{M}{4}, \bbt_i + \frac{M}{4}]^2$. 
Under fading conditions, the channel between a transmitter $i$ and any receiver $r(j)$, at any scheduling instant, is composed of two components $H_{ij} = H_{ij}^P H_{ij}^F$, where the path gain is given by $H_{ij}^P = \lVert \mathbf{t}_i - \mathbf{r}_j \rVert ^ {-2.2}$ and the fading coefficient is randomly drawn $H_{ij}^F \sim \text{Rayleigh}(1)$. These parameter values are approximately representative of realistic path loss and fading in several actual wireless environments.
%We fix $H_{ij}^P$ and sample $H_{ij}^F$ at each power allocation instant. 
In Section~\ref{Ss:performance_comparison} we assume that the underlying topology of the network is fixed whereas in Section~\ref{Ss:spat_den} we study UWMMSE's generalization when the size or density of the network changes.

The UWMMSE architecture used is composed of $4$ unrolled WMMSE layers with each layer $k$ having two $2$-layered GCNs modeling the function $\Psi$ in~\eqref{E:unfold_1}.
The hidden layer dimension of all GCNs is set to $5$. 
$10000$ training iterations are performed per epoch with a maximum of $20$ epochs. 
Batch size is fixed at $64$ and learning rate is set to \num{1e-3}.
We run $100$ test iterations with the same batch size. 
Unless otherwise specified, the network size $M$ is fixed at $20$.

\subsection{Performance comparison}\label{Ss:performance_comparison}

We compare the performance attained by UWMMSE with that of established baselines in the challenging low-noise regime [$\sigma = \num{2.6e-5}$ in~\eqref{E:data_rate}] where the achievable capacity is highly determined by the interference between users.
We choose the following prior-arts for comparison:

\begin{enumerate}
    \item \textit{WMMSE} \cite{shi2011iteratively} forms the baseline for our experiments. We set a maximum of $100$ iterations per sample.
    \item \textit{Truncated WMMSE} (Tr-WMMSE) provides a performance lower bound to UWMMSE.  
    We fix the number of iterations to $4$ to match UWMMSE unrollings.   
    \item \textit{MLP} \cite{sun2018learning} is employed in a supervised setting to replicate WMMSE output.
    \item \textit{REGNN} \cite{eisen2020optimal} addresses the specific problem of \textit{binary} power allocation in Rayleigh interference channels.  
    \item \textit{IGCNet} \cite{shen2019graph} addresses the power allocation problem in Gaussian interference channels.  

\end{enumerate}

%\textcolor{blue}{[question: Did we re-train each of the other models for the particular parameter values of Rayleigh fading, noise std dev, etc used in this paper? or did we use the trained models of those other approaches e.g. IGCNet as is?]}

The comparisons are shown in Fig~\ref{Fig:performance_comparison}(a). 
As channel states are sampled randomly, there can be significant variation in the utility value for individual samples depending on the respective interference patterns, even under optimal power allocation.
The figure reveals that UWMMSE matches the performance of WMMSE. 
It is also interesting to note that UWMMSE bridges the gap between WMMSE and its truncated version owing to the additional flexibility afforded by learnable parameters. MLP, which learns a functional mapping of the WMMSE output, beats Tr-WMMSE but still falls short of WMMSE by a significant margin. 
This shows that supervised methods are limited by the quality of their training signals. 
% and often do not generalize to out-of-sample data. 
%\textcolor{blue}{[I did not fully understand this statement; are we saying this because MLP was trained on channel data of a different distribution than the one used here?]}
On the other hand, both REGNN and IGCNet -- originally designed and tested in high-noise regimes -- prove to be inadequate to match the performance of WMMSE in the more challenging low-noise setting. 

In addition to achieving a sum-rate that is close to optimal, it is essential to minimize the time taken for power allocation as the channel states tend to change rapidly. 
To that end, we provide a computation time comparison\footnote{All computations were performed on an Nvidia Quadro T2000 GPU.} in Table~\ref{tab:performance}.  UWMMSE, which takes close to $2$ milliseconds (ms) per sample, is significantly faster than WMMSE which takes around $16$ ms per sample. 
All the other learning-based methods~ \cite{shen2019graph,sun2018learning,eisen2020optimal} have a processing time similar to that of UWMMSE, however, none of them achieves the same performance, which is the main advantage of our method over existing algorithms. 

\subsection{Generalization to variations in network density and size}\label{Ss:spat_den}

We consider two scenarios in which a wireless network undergoes variations in terms of the density and size of the underlying topology. 
We simulate a dynamic topology by varying the spatial density of the static network from the previous experiment. 
A density factor $d$ is used as a control parameter for this experiment.
More precisely, for each value of $d$, a transmitter is dropped at location $\mathbf{t}_i^d = {\bbt_i}/{d}$. 
Its paired receiver $r(i)$, however, is still dropped uniformly at random at location $\mathbf{r}^d_i \in [\bbt_i^d - \frac{M}{4}, \bbt_i^d + \frac{M}{4}]^2$. 
Effectively, the spatial density is varied according to relative positions of the transmitters while maintaining a degree of stochasticity in the receiver positions. 
We focus on networks that are denser than the original one ($1.0 \leq d \leq 5.0$) to analyze the challenging case of increasing interference due to geographical proximity; see Fig.~\ref{Fig:performance_comparison}(b).

\begin{table}[t]
\centering
\caption{Performance and time comparisons of all methods}
\vspace{2mm}
\begin{tabular}{l  c  c  c  c  c}
\hline
%Algorithm  & Train time (min)  & Mean sum-rate & Test time (ms) \\
Algorithm  & Train  & Mean & Test \\
& time (min) & sum-rate & time (ms) \\
\hline
WMMSE~\cite{shi2011iteratively} &  -  &  82.94  & 16.0 \\ 
%\hline
Tr-WMMSE &  -  &   76.49  & 1.0 \\
%\hline
MLP~\cite{sun2018learning} & 0.5 &  78.17   &  3.2 \\
%\hline
REGNN~\cite{eisen2020optimal} &  15  &  57.92   & 2.5  \\
%\hline 
IGCNet~\cite{shen2019graph} &  5   &  55.30   & 3.0  \\
%\hline
UWMMSE & 15 &  \textbf{83.21}  & \textbf{2.0} \\
\hline
\end{tabular}
\label{tab:performance}
%\textcolor{blue}{[do we intend mean sum-rate instead of sum-rate?]}
\end{table}

In this experimental setup, we compare UWMMSE trained on the static network against WMMSE and Tr-WMMSE as baselines. Clearly, there is an approximately constant but moderate gap in performance between WMMSE and UWMMSE for all density values except $d=1.0$, as UWMMSE achieves limited generalization to unseen network topologies. 
As an improvement, we introduce a robust version of UWMMSE (Ro-UWMMSE), that is trained on multiple network topologies with varying spatial density $d \in [0.5,5.0]$. 
As evident in Fig.~\ref{Fig:performance_comparison}(b), Ro-UWMMSE follows WMMSE performance closely and, importantly, improves upon Tr-WMMSE demonstrating the value of learning even for varying densities.

We now consider a variable-size setup that involves random insertion or deletion of nodes in a wireless network. 
To that end, a set of transceivers is either removed from the network or a new set of transceivers are added to the network at every scheduling instant. 
Note that new transceivers are still added in the interval $\bbt_i \in [-M, M]^2$ for all $i > M$, where $M$ is the original network size, to avoid any expansion or contraction of the overall area of the multi-sized topology. 
Corresponding receivers are dropped $\bbr_i \in [\bbt_i - \frac{M}{4}, \bbt_i + \frac{M}{4}]^2$. 
Fading coefficients are sampled independently for each individual topology.
We evaluate model performance on networks of size $N \in \{10, \ldots,30\}$ by either removing nodes from the original network ($N < M$) or adding new unseen nodes to it ($N>M$), as shown in Fig.~\ref{Fig:performance_comparison}(c). 

Similar to the previous experiment, we compare our method against WMMSE and Tr-WMMSE as baselines and, as expected, UWMMSE performs best on networks that are of the same size as the training samples. To make UWMMSE robust against these variations, we train a robust version Ro-UWMMSE on networks of multiple sizes by randomly generating a batch of networks of size $M \in \{10, \ldots, 30\}$ at each training step. As evident in Fig.~~\ref{Fig:performance_comparison}(c), Ro-UWMMSE is able to maintain performance that is close to WMMSE and, thus, illustrates the generalization capacity of our methodology.

%%%%%%%%%%%%%%%%%%%%%%%%%%%%%%%%%%%%%%%%%%
\section{Conclusions}\label{S:Conclusions}
%%%%%%%%%%%%%%%%%%%%%%%%%%%%%%%%%%%%%%%%%%

We proposed UWMMSE, a novel neural network based approach to solve the problem of power allocation in wireless networks.
The layered architecture of the presented method was derived from the algorithmic unfolding of the classical WMMSE algorithm, thus, UWMMSE naturally incorporates domain-specific elements augmented by trainable components.
These components are parameterized by GNNs to account for and leverage the inherent graph representation of communication networks. We have demonstrated that UWMMSE achieves performance comparable to that of WMMSE while being significantly faster than it. Current efforts include analyzing -- theoretically and empirically -- the model performance under missing, noisy, or even adversarial channel information. 

\newpage

% \section{REFERENCES}
% \label{sec:refs}

\bibliographystyle{IEEEbib}
\bibliography{main}

\end{document}